# An Integrated Bell-State Analyzer on a Thin Film Lithium Niobate Platform

Uday Saha[1,2] and Edo Waks[1,2,3]*

[1]Department of Electrical and Computer Engineering, University of Maryland, College Park, Maryland, USA, 20742

[2]Institute for Research in Electronic and Applied Physics (IREAP), University of Maryland, College Park, Maryland, USA, 20742

[3]Joint Quantum Institute (JQI), University of Maryland, College Park, Maryland, USA, 20742

*Corresponding Author: edowaks@umd.edu

## Abstract

Trapped ions are excellent candidates for quantum computing and quantum networks because of their long coherence times, ability to generate entangled photons as well as high fidelity single- and two-qubit gates. To scale up trapped ion quantum computing, we need a Bell-state analyzer on a reconfigurable platform that can herald high fidelity entanglement between ions. In this work, we design a photonic Bell-state analyzer on a reconfigurable thin film lithium niobate platform for polarization-encoded qubits. We optimize the device to achieve high fidelity entanglement between two trapped ions and find >99% fidelity. The proposed device can scale up trapped ion quantum computing as well as other optically active spin qubits, such as color centers in diamond, quantum dots, and rare-earth ions.

Trapped ions are one of the most advanced platforms for quantum computing and quantum networks. They exhibit long coherence times [1–3], naturally emit photons entangled with their internal qubit memories [4,5], and support high-fidelity single- and two-qubit gates [6,7]. But quantum computers and networks based on trapped ions also require photonic devices in order to interconnect different nodes of the network [8,9]. The key optical component in these devices is the photonic Bell-state analyzer [10], which heralds the successful entanglement between distant qubits. Such entanglement provides a photonic interconnect between trapped ions, which can be used to scale up trapped ion quantum computers [2,11] as well as enable long-distance quantum networks [8,12–14]. But to realize these interconnects in a scalable way requires compact reconfigurable chip-integrated devices that can perform Bell-state analysis between trapped ions at different nodes [8,12] .

Thin film lithium niobate is a promising material platform to implement photonic Bell analyzers on a compact chip. Lithium niobate has a broad transparency window over the visible spectrum [15], making it compatible with many ion species. It also possesses a high electro-optic coefficient, which enables reconfigurable photonic devices with ultra-fast switching speed [16,17]. Recent progress in fabrication has enabled ultra-low loss thin film lithium niobate waveguides in both the visible [15] and telecom range [18]. But applying thin-film lithium niobate devices to photons generated by trapped ions is challenging because trapped ions naturally emit polarization-encoded photonic qubits [19,20]. Thin-film lithium niobate waveguides possess small sidewall angles [21] and require partial etching to achieve low waveguide loss [18,22]. These properties create large polarization anisotropy, which makes it difficult to engineer a photonic Bell-state analyzer for polarization qubits. Although in principle it is possible to convert polarization qubits to time-bin qubits, the relatively long excited-state lifetimes of trapped ions [23] would necessitate impractically large on-chip delay lines. Realizing a photonic Bell-state analyzer with polarization-encoded qubits in thin-film lithium niobate would therefore be an important step towards applying this powerful platform to trapped ion quantum interconnects.

In this work, we propose and analyze a design for a photonic Bell-state analyzer for polarization-encoded qubits in thin-film lithium niobate. To overcome the polarization anisotropy, we optimize the dimensions of the device to achieve the desired unitary operation of the transverse electric (TE) and transverse magnetic (TM) polarization components simultaneously. We design the Bell-state analyzer to operate at the main transition wavelength (493.55 nm) of trapped barium ions ($^{133}Ba^+$ and $^{138}Ba^+$), which are a leading candidate for optically active qubits for trapped ion quantum computing [2,24]. We further optimize the input coupler to achieve nearly identical light coupling efficiencies for the TE and TM modes (66.16% and 65.52%, respectively). Our numerical analysis shows the optimized device can mediate entanglement between distant ions with fidelities exceeding 99%. The proposed Bell-state analyzer provides a promising approach to interconnect trapped ions, which would enable the scaling of trapped ion quantum computers [2,11] as well as other optically active spin qubits such as color centers in diamond [25], quantum dots [26], and rare-earth ions [27].

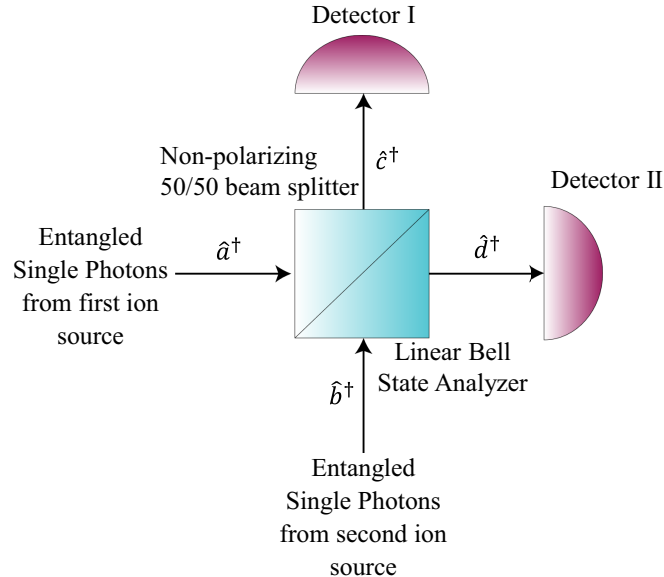

**Fig. 1** Entanglement experiment of two ions through quantum interference on a nonpolarizing 50/50 beam splitter.

Fig. 1 illustrates the operating principle of the photonic Bell-state analyzer and how it mediates entanglement between trapped ions. The Bell-state analyzer is composed of a non-polarizing 50/50 beam splitter and photon detectors. Each ion emits a spin-photon with an entangled quantum state given by $|\psi\rangle = |\uparrow\rangle |V\rangle + |\downarrow\rangle |H\rangle$, where $|\uparrow\rangle$ and $|\downarrow\rangle$ represent the spin up and spin down states of the ions, and $|H\rangle$ and $V\rangle$ denote two orthogonal polarization states of the emitted single photons. The photon from each ion is injected into the respective input port of the beam splitter. A coincidence detection event between the two detectors heralds a projective measurement of the two-photon state onto the Bell-state $|\psi^-\rangle_p = (|H\rangle |V\rangle - |V\rangle |H\rangle)/\sqrt{2}$ [5]. If no coincidence is detected the measurement produces an inconclusive result. Although this measurement is probabilistic, it is sufficient to herald entanglement between ions in order to scale up trapped ion quantum computers [2,11] or establish a quantum network [12]. Improved configurations can yield higher efficiencies [28], but in this work we will consider this simple Bell analyzer device.

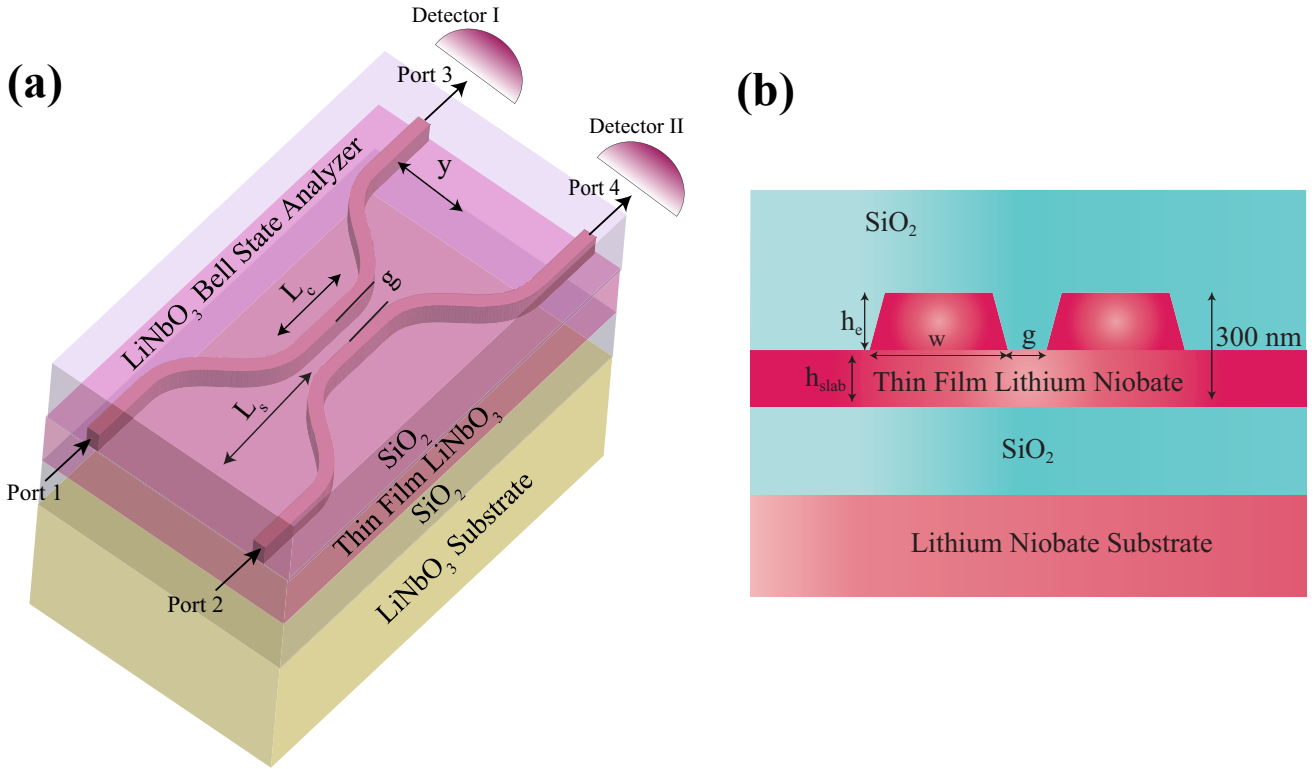

**Fig. 2** (a) 3D schematic of the proposed non-polarizing on-chip beam splitter fabricated on a thin-film lithium niobate platform. (b) Cross-sectional view of the Bell-state analyzer in the coupling region.

Although conceptually simple, chip-integration of this device is challenging due to the requirement that the beam splitter must be non-polarizing. To satisfy this requirement, we consider the directional coupler structure shown in Fig. 2(a), which acts as an on-chip beam splitter. The directional coupler is composed of two waveguides that are brought together at a coupling region of length $L_c$, where they can interact through their evanescent fields. Each waveguide supports a TE and TM mode, which define the polarization qubit basis of a propagating photon. The two input modes enter through ports 1 and 2, and the two output modes leave through ports 3 and 4, as illustrated in Fig. 2(a). Photon detectors at the output ports herald entanglement generation. To connect these ports to the coupling region, we use S bends that taper the initial distance ($y$) of 2 μm between the waveguides over a distance of $L_s$. Fig. 2(b) shows the cross-sectional view of the device structure in the coupling region. The layer structure is composed of 300 nm thick X-cut lithium niobate on 2 μm thick silicon dioxide ($SiO_2$). Each waveguide is formed from a partially etched rib waveguide structure with 75° side angles [21] and partial etch depth and width of $h_e$ and $w$, respectively. We define the slab waveguide thickness as $h_{slab} = (300 - h_e)$ nm. The waveguides are separated by a gap distance of $g$. They are also encapsulated in a 2 μm thick $SiO_2$ cladding layer.

The goal is to design a beam splitter that exhibits 50/50 power splitting ratios for both TE and TM modes. To gain a better understanding of how to achieve this condition, we apply coupled mode theory to analyze the coupling region of the device [29]. The coupled waveguides form two new modes, a symmetric and anti-symmetric mode (Fig. S1 in the Supporting Information). The power splitting between the waveguides depends sinusoidally on the difference of the effective refractive indexes of these modes ($\Delta n$), which we define as the coupling strength. If light is initially injected into port 1 with power $P_I$, then the power in ports 3 and 4 are given by $P_3 = P_I \cos^2\left(\frac{\pi \Delta n}{\lambda_0} L_c\right)$ and $P_4 = P_I \sin^2\left(\frac{\pi \Delta n}{\lambda_0} L_c\right)$, where $\lambda_0$ is the operating wavelength.

Due to the large anisotropy in the shape of the waveguides, the TE and TM modes will in general have different $\Delta n$, which will lead to different power splitting ratios for the two polarizations. To achieve a Bell-state analyzer, we need to first minimize this disparity. We denote $\Delta n_{TE}$ and $\Delta n_{TM}$ as the coupling strength for the TE and TM modes, respectively. We define the ratio $\xi = \frac{\Delta n_{TE}}{\Delta n_{TM}}$ as the figure of merit for our device, with the optimal device achieving $\xi = 1$.

We perform all simulations using a three-dimensional eigenmode expansion solver (Lumerical Mode solutions). In all calculations, we set the refractive indexes of lithium niobate and $SiO_2$ to be 2.34 [30,31] and 1.462 [32], respectively.

To achieve $\xi = 1$, we vary the width and etch depth of the Bell-state analyzer with a gap distance of $g = 65$ nm (Fig. 3 (a)). The contour line of $\xi = 1$ passes through multiple combinations of the width and etch depth. From these combinations, we select a width of 475 nm and etch depth of 110 nm where $\xi = 1$. Then, we observe the effect of the gap on $\xi$ at that width and etch depth (Fig. 3(b)). We find that $\xi$ increases monotonically with the gap of the coupling region. We achieve $\xi = 1$ at our previous simulated gap of 65 nm. There are multiple combinations of the width, etch depth, and gap that can lead to $\xi = 1$ and polarization insensitivity for the coupling region of the Bell-state analyzer. We select a combination of $w = 475$ nm, $h_e = 110$ nm, and $g = 65$ nm for our following analysis where $\xi = 1$.

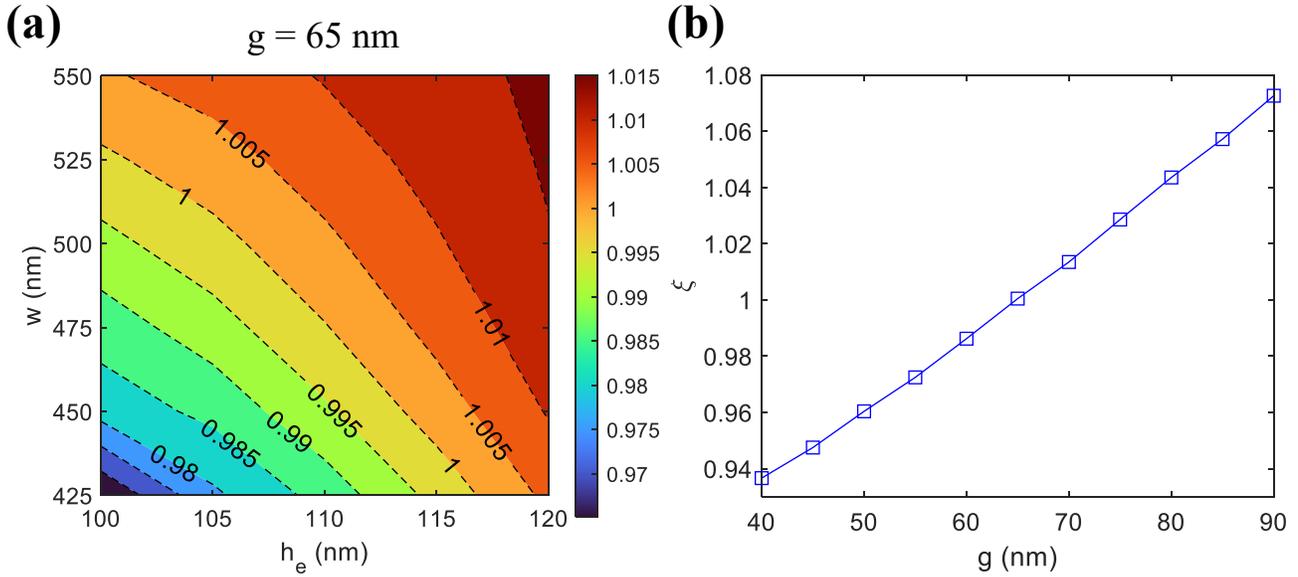

**Fig. 3** The change in $\xi\,(=\Delta n_{TE}/\Delta n_{TM})$ as a function of (a) the width ($w$) and etch depth ($h_e$) of the Bell-state analyzer at $g = 65$ nm and (b) as a function of the gap distance ($g$) at $w = 475$ nm and $h_e = 110$ nm.

In order to design the full device, we must also consider the S bend regions (Fig. 2 (a)), which can significantly contribute to the power splitting ratio. The bending regions create additional complications by introducing greater anisotropy as well as polarization-dependent loss. Using the previously optimized width and etch depth, we optimize the bending length ($L_s$) to achieve minimal bending losses and a small device footprint that can route quantum information efficiently (Fig. S2 in the Supporting Information). We choose $L_s$ to be 30 microns, as this value results in high transmission for both the TE and TM modes (0.993 and 0.994, respectively).

Using the previously optimized width ($w$), etch depth ($h_e$), gap ($g$), and bending length ($L_s$), we then simulate the complete device with the S-bends. To analyze the polarization-insensitive operation of the device, we investigate how the normalized power values at ports 3 and 4 change with the coupling length ($L_c$) for the TE and TM modes (Fig. 4(a)). In this case, the normalized power at both ports do not match for the TE and TM modes, resulting in a polarization-dependent power splitting ratio for the Bell-state analyzer. This problem arises because the S-bend region contributes significantly to the power coupling between the waveguides. This coupling is highly sensitive to polarization, leading to polarization-dependent power splitting for the overall device. Given the effect of the S-bend, we need to reoptimize the parameters to achieve a polarization-independent power splitting ratio for the whole device.

To achieve polarization insensitivity for the full device including the S-bends, we vary the gap, which significantly controls the power splitting ratio of the Bell-state analyzer, while keeping the other parameters fixed at their previously optimized values ($w = 475$ nm, $h_e = 110$ nm, $L_s = 30$ μm). We define a polarization-dependence parameter, $\zeta$, which is the difference in the normalized powers of the TE and TM modes of port 3. To achieve any arbitrary polarization insensitive power splitting ratio, we calculate the root mean square value of $\zeta$ for coupling lengths of 1 to 25 μm, which contain all the possible power splitting ratios, and define the parameter as $\delta$. We then minimize $\delta$ by varying the gap, which occurs at $g = 40$ nm (Fig. 4(b)). Fig. 4(c) shows the normalized power values at ports 3 and 4 for the TE and TM modes with respect to the coupling length ($L_c$) at $g = 40$ nm, which virtually overlap. As a result, we can approximately achieve any polarization insensitive power splitting ratio by varying the coupling length from 6.4 μm to 21.5 μm. These results show that in addition to achieving a 50/50 power splitting ratio, we can attain any desired ratio by varying the coupling length. This ability is key to realizing different types of photonic quantum gates for polarization-based qubits [33,34].

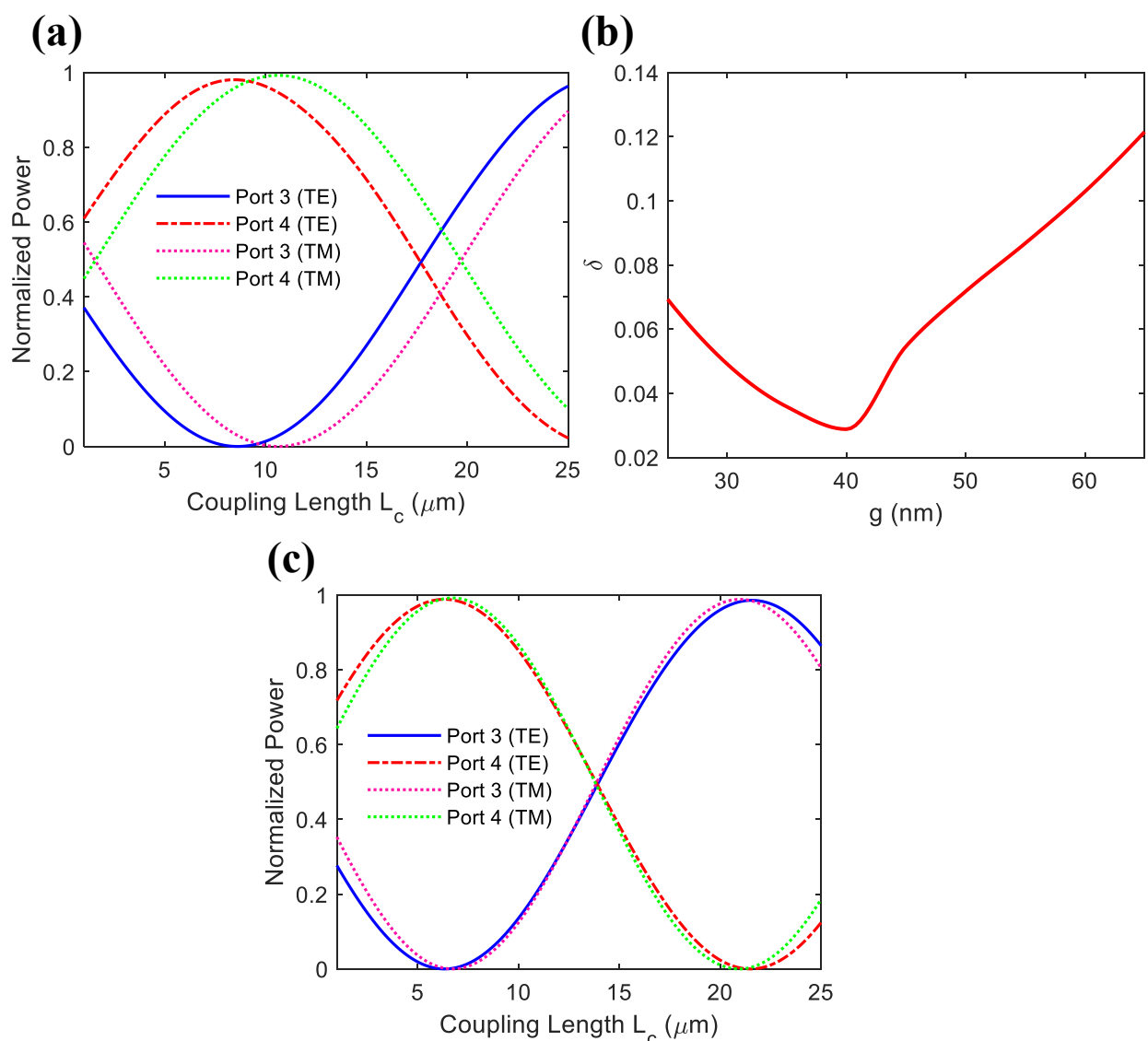

**Fig. 4** (a) The change in the normalized power at ports 3 and 4 with respect to the coupling length ($L_c$) for TE and TM polarization at $g = 65$ nm. (b) The variation of $\delta$ as a function of $g$. (c) The change in the normalized power at ports 3 and 4 as a function of the coupling length ($L_c$) for TE and TM polarization at $g = 40$ nm.

To predict our device performance for entangling two remote ions, we calculate the error in entanglement. On a Bell-state analyzer, the input-output relationship can be written in the following way:

$$\begin{bmatrix} \hat{a}^\dagger_{h(v)} \\ \hat{b}^\dagger_{h(v)} \end{bmatrix} = \begin{bmatrix} t_{h(v)} & r_{h(v)} \\ r_{h(v)} & -t_{h(v)} \end{bmatrix} \begin{bmatrix} \hat{c}^\dagger_{h(v)} \\ \hat{d}^\dagger_{h(v)} \end{bmatrix} \quad (1)$$

in which $\hat{a}^\dagger_{h(v)}$ and $\hat{b}^\dagger_{h(v)}$ represent the input bosonic operators of ports 1 and 2 for horizontal (vertical) polarization and $\hat{c}^\dagger_{h(v)}$ and $\hat{d}^\dagger_{h(v)}$ are the output bosonic operators of ports 3 and 4 for horizontal (vertical) polarization. The parameters $t_{h(v)}$ and $r_{h(v)}$ are the transmission and reflection coefficients of the photonic Bell-state analyzer for horizontal (vertical) polarization, respectively. Upon simultaneous detection of photons in the detectors, we can write the detected state as

$$|\psi\rangle_{detected} = \frac{1}{r_h^2 r_v^2 + t_h^2 t_v^2}(r_h r_v |\downarrow\rangle_a|\uparrow\rangle_b - t_h t_v |\uparrow\rangle_a|\downarrow\rangle_b) \quad (2)$$

The fidelity of detecting the entangled ion state $|\psi^-\rangle_I = (|\downarrow\rangle_a|\uparrow\rangle_b - |\uparrow\rangle_a|\downarrow\rangle_b)/\sqrt{2}$ will be

$$F = \|\langle\psi_I^-|\psi_{detected}\rangle\|^2 = \frac{1}{2}\frac{(r_h r_v + t_h t_v)^2}{r_h^2 r_v^2 + t_h^2 t_v^2} \quad (3)$$

Then, we define the entanglement error as

$$E = 1 - F \quad (4)$$

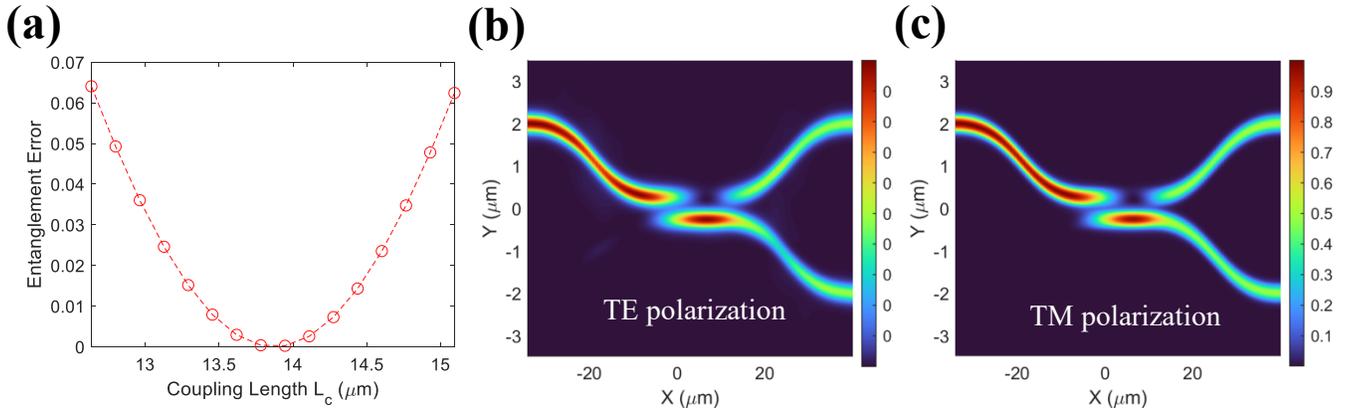

**Fig. 5** (a) Variation of the error of entangling two ions as a function of $L_c$ (from 12.5 μm to 15 μm). (b) and (c) represent the electric field propagation through the Bell-state analyzer at $L_c$ = 13.95 μm for TE and TM polarization, respectively.

Fig. 5(a) depicts the error of the entanglement of the two ions as a function of the Bell-state analyzer's coupling length ($L_c$). To obtain high fidelity entanglement between two remote ions, we simulate a sweep of the coupling length from 12.5 μm to 15 μm, where a 50/50 power splitting ratio occurs, as shown in Fig. 4(c). We achieve a minimum error of 2.48 ×10$^{-4}$ at $L_c$ = 13.95 μm. This corresponds to a power splitting for ports 3 and 4 of 49.7/48.9 (Fig. 5(b) and 50.7/48.3 (Fig. 5(c)) for the TE and TM modes, respectively. These findings demonstrate we can achieve a compact integrated Bell-state analyzer on a thin film lithium niobate platform to generate high fidelity entanglement between remote qubit systems.

Although we design this polarization-insensitive photonic Bell-state analyzer for remote ion entanglement, coupling light into the chip independent of polarization is crucial to maintain high fidelity entanglement between ions. If the TE and TM modes couple with different efficiencies, that will destroy the polarization insensitivity of the device and reduce the fidelity of entanglement. Therefore, we analyze and optimize the coupling efficiencies for the TE and TM modes.

To couple light into the chip, we consider a specific coupling scheme based on a lensed fiber. This approach can maintain the polarization insensitivity and yield high coupling efficiencies [35]. In this scheme, a lensed fiber is edge-coupled to the chip and the fiber numerical aperture (NA) is matched to waveguide's NA. We determine the coupling efficiencies by calculating the mode overlap integral between the lensed fiber mode and waveguide mode (Fig. 6 (a) and (b) represent the TE and TM modes, respectively).

Fig. 6 (c) represents the light coupling efficiency for the TE and TM modes with respect to the NA of the lensed fiber. At lower NA, the gaussian beam of the lensed fiber is larger than the waveguide mode. Due to the mode mismatch, we achieve lower coupling efficiencies at lower NAs. By increasing the NA, we can decrease the mode size of the lensed fiber and make it closer to the waveguide mode to achieve higher coupling efficiencies. However, we vary up to 0.6 NA which is commercially available. At this NA value, we obtain 66.16% and 65.52% coupling efficiencies for TE and TM polarization, respectively. Taking these coupling efficiencies into account, we further calculated the error in entanglement and found an error of 2.54 ×10$^{-4}$ in entangling two remote ions. Through our design, we can incorporate light in a polarization-independent manner, which enables an efficient interface for transferring quantum information from polarization qubits.

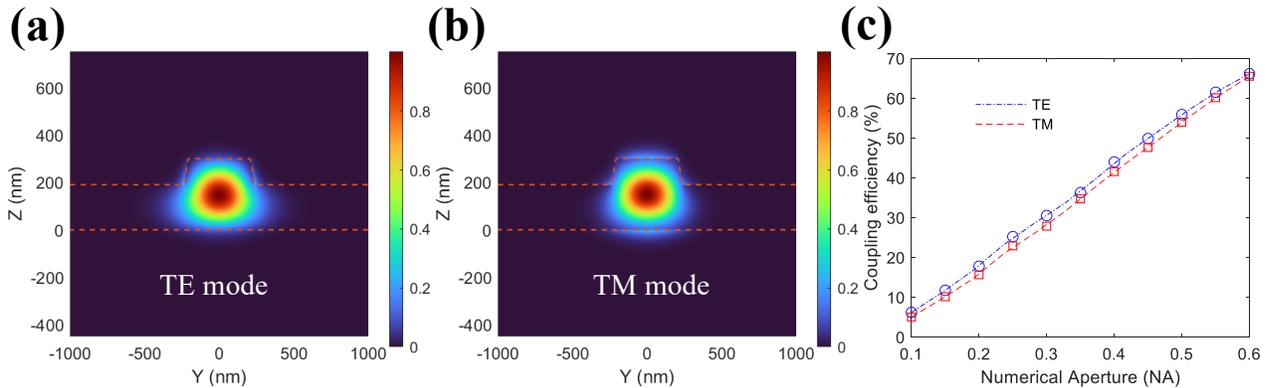

**Fig. 6** (a) and (b) represent the electric field intensity of the waveguide for TE and TM polarization, respectively. (c) The change in the light coupling efficiency from fiber to chip with respect to the numerical aperture (NA) of the lensed fiber.

In summary, we design an integrated photonic Bell-state analyzer on a thin-film lithium niobate platform that exhibits greater than 99% fidelity for the entanglement of two remote trapped ion qubit systems. To overcome polarization-mode anisotropy, we optimize the etch depth, width, gap, bending length, and coupling length of the Bell-state analyzer and achieve polarization independent 50/50 power splitting and minimum error in entangling two remote ions. We also analyze and optimize the numerical aperture of a lensed fiber to achieve a polarization-insensitive light coupling interface and achieve identical coupling efficiencies for TE and TM polarization (66.16% and 65.52%, respectively). Apart from entangling two remote trapped ions, we can achieve any polarization-independent power splitting ratios in the directional coupler by adjusting the coupling length, which can be used to realize different quantum gates. Moreover, the design of the photonic Bell-state analyzer can be modified to create optical interconnects between optically active spin qubit systems, such as color centers in diamond, quantum dots, and rare earth ions, etc. By using electro-optic modulation, our proposed device can be engineered to act as a reconfigurable polarization-maintaining switch fabric, which may be useful for other quantum applications. Therefore, this integrated photonic Bell-state analyzer design could enable a new generation of compact and reconfigurable integrated photonic devices that can serve as efficient quantum interconnects for quantum computers and sensors.

## Note

The authors declare no competing financial interest.